\documentclass[titlepage,twoside,10pt,letter]{article}
\usepackage[margin=1.2in]{geometry}
\usepackage{enumerate}
\usepackage{times}
\usepackage{amsmath}
\usepackage{amsfonts}
\usepackage{amssymb}
\usepackage{graphicx}
\usepackage{mathrsfs}
\usepackage{tabularx}
\usepackage{caption}
\usepackage{bbold}
\usepackage{color}
\usepackage{fancybox}
\usepackage{verbatim}
\usepackage{hyperref}
\usepackage{tikz}
\usepackage{float}
\usetikzlibrary{arrows,automata}
\usetikzlibrary{trees}

\def\b1{\mathbb{1}}

\def\cB{\mathcal{B}}

\def\e1{\mathsf{1}}

\def\of0{(0)}

\def\d{\partial}

\def\bf0{\mathbf{0}}
\def\cp1{\mathbb{CP}^1}

\begin{document}
	
\title{\textbf{{Bartlett's delta in the SABR model}}}
\author{\textbf{Patrick S. Hagan}\\
Gorilla Science\\
PatHagan@GorillaSci.Com
\and \textbf{Andrew Lesniewski}\\
Department of Mathematics\\
Baruch College\\
One Bernard Baruch Way\\
New York, NY 10010\\
USA}
\date{First draft: April 14, 2016\\
This draft: \today}
\maketitle

\begin{abstract}
The presence of stochastic volatility in an option model impacts the values of the hedge ratios (the ``greeks''), and in particular the option delta. In the context of the SABR model, the greeks were calculated in \cite{hklw02} based on the asymptotic expression for the implied volatility derived there. In \cite{b06}, the option delta of \cite{hklw02} was modified to take into account the effects of the correlation between the dynamics of the forward and the stochastic volatility. It was empirically observed there that the modified delta (``Bartlett's delta'') provides a more accurate and robust hedging strategy than the original SABR delta. In this paper we refine the analysis of hedging strategies carried out in \cite{b06}. In particular, we provide a justification of the empirical observations regarding the robustness of the modified delta. This is done by means of an asymptotic analysis of the explicit expression for the implied volatility derived in \cite{hklw02}. In particular, we show that the modified option delta is practically insensitive to the choice of the the CEV parameter $\beta$.
\end{abstract}

\section{\label{introSec}Introduction}

The key requirement of an option model, in addition to its utility as an accurate pricing tool, is its ability to produce reliable risk metrics. This allows the portfolio manager or market maker to confidently put on appropriate hedges on his options positions in a way that reflects his view or mandate.

The presence of stochastic volatility in an option model impacts the values of the the option greeks, and in particular the option delta. In this note we are concerned with hedging under the SABR model of volatility smile (\cite{hklw02}, \cite{hlw05}, \cite{hklw14}). Originally, the values of the greeks under the SABR model were calculated in \cite{hklw02} based on the asymptotic expression for the implied volatility. In \cite{b06}, the option delta proposed in \cite{hklw02} was modified to take into account the effects of the correlation between the dynamics of the forward and the stochastic volatility. It was empirically observed there that the modified delta provides a more accurate and robust hedging strategy than the original SABR delta.

The results described below are a refinement of the work presented in \cite{b06}. The SABR model's specification requires four parameters $\sigma,\alpha,\beta,\rho$, whose values are calibrated to options market prices. According to the prevailing market practice, one of these parameters, the CEV exponent $\beta$ is usually set to a pre-specified value, while the remaining three parameters are optimized. This practice is justified by the fact that there is a degree of redundancy between the CEV exponent $\beta$ and the correlation parameter $\rho$ in the SABR parameterization of the smile curve. While this choice introduces a higher degree of stability of the model parameters, it brings up the question whether the resulting hedges are equally robust, i.e. whether the choice of $\beta$ made by the portfolio manager has a significant impact on the hedging strategy. 

It was argued in \cite{b06} that the modified delta $\Delta^{\mathrm{mod}}$ introduced there leads to more robust hedges than the classic SABR delta \cite{hklw02}, namely, across a large range of strikes, it is nearly independent of the choice of $\beta$. This claim was supported there by empirical and numerical arguments, see also \cite{am10}, \cite{hklw14}, and \cite{hw16}. The purpose of this note is to provide a theoretical justification of this claim. This is done by means of an asymptotic analysis of the explicit expression for the implied volatility derived in \cite{hklw02}. Furthermore, we show here that the modified delta of an at the money option is exactly independent of the choice of $\beta$. 

The robustness of $\Delta^{\mathrm{mod}}$ is of direct and measurable practical significance. Proper hedging allows the portfolio manager or market maker better implement his views, which may have an impact on his P\&L. Accurate hedge ratios allow for reliable portfolio return attribution, which facilitates his communication with the management, clients, and auditors. Also, from the perspective of regulatory requirements and model risk management, the advantage of the modified SABR delta is clear. It provides a robust option delta, which is insensitive to possible model misspecification, and it thus is a model risk mitigant.

We consider a European call or put struck at $K$ and expiring in $\tau$ years from the current time, and let $F$ denote the current value of the underlying forward. The implied volatility curve is a function $\sigma^{\mathrm{imp}}=\sigma^{\mathrm{imp}}(\tau,F,K,\sigma)$ such that when combined with the Black-Scholes formula, it yields (close approximations to) the market option prices. Two market observable quantities are of particular interest to option traders: the at the money implied volatility,
\begin{equation}
\sigma^{\mathrm{ATM}}=\sigma^{\mathrm{imp}}(\tau,F,F,\sigma),
\end{equation}
and the skew,
\begin{equation}
\eta=\frac{\d \sigma^{\mathrm{imp}}(\tau,F,K,\sigma)}{\d K}\,\Big|_{K=F}\,.
\end{equation}
The latter is the slope of the volatility curve calculated at the money. These two quantities are model independent, and can be directly inferred from option prices. Any reasonable volatility smile model, regardless of its specification, can be calibrated so that these two quantities match the market values sufficiently closely.

Our main results is that, for each strike $K$, the modified SABR delta $\Delta^{\mathrm{mod}}$ has approximately the following structure:
\begin{equation}
\Delta^{\mathrm{mod}}=\Delta^{\mathrm{Black-Scholes}}+\mathrm{Vega}^{\mathrm{Black-Scholes}}\times\eta.
\end{equation}
In other words, other than the standard Black-Scholes greeks calculated for strike $K$, the modified SABR delta does not involve any details of the smile model specification. In contrast, the standard SABR delta has the structure
\begin{equation}
\Delta=\Delta^{\mathrm{Black-Scholes}}+\mathrm{Vega}^{\mathrm{Black-Scholes}}\times(\eta+\text{model dependent term}).
\end{equation}
The last term in the expression above is responsible for potential mishedging in case of model miscalibration discussed in \cite{b06}.

\section{\label{sabrSec} The SABR model}

The dynamics of the SABR model of option implied volatility is specified in terms of two state variables: the forward $F_t$ and the instantaneous volatility $\sigma_t$. Explicitly, the dynamics is given by the system of stochastic differential equations:
\begin{equation}
\begin{split}
dF_t&=\sigma_t C(F_t)dW_t,\\
d\sigma_t&=\alpha\sigma_t dZ_t,
\end{split}
\end{equation}
where $W_t$ and $Z_t$ are two Brownian motions with
\begin{equation}
dW_t dZ_t=\rho dt.
\end{equation}
The positive function $C(F)$ determines the backbone of the volatility smile, and is usually assumed to be of the CEV form
\begin{equation}
C(F)=F^\beta,
\end{equation} where $\beta\leq 1$ is the CEV parameter\footnote{In order to handle negative forward rates in interest rate markets, some practitioners choose $C(F)=(F+\theta)^\beta$, with $\theta>0$.}. This will be our default choice in the following.

The normal implied volatility in the SABR model is given by the following asymptotic expression \cite{hklw02} in the (small) parameter $\varepsilon=\alpha^2\tau$:
\begin{equation}\label{implVol}
\sigma^{\mathrm{imp}}=\alpha\;
\frac{F-K}{D(\zeta)}\;
\Big\{1+\Gamma\varepsilon+O(\varepsilon^2)\Big\},
\end{equation}
where $F$ denotes here the currently observed value of the forward. The distance function $D(\zeta)$ entering the formula above is given by
\begin{equation}
D(\zeta)=\log\Big(\frac{I(\zeta)+\zeta-\rho}{1-\rho}\Big),
\end{equation}
where
\begin{equation}
I(\zeta)=\sqrt{1-2\rho\zeta+\zeta^2}\,,
\end{equation}
and where
\begin{equation}
\begin{split}
\zeta&=\frac{\alpha}{\sigma}\;\int_K^{F}\frac{dx}{C(x)}\\
&=\frac{\alpha}{\sigma}\,\frac{F^{1-\beta}-K^{1-\beta}}{1-\beta}\,.
\end{split}
\end{equation}
The parameter $\sigma$ denotes the currently observed value of the instantaneous volatility.

Various forms of the first order correction $\Gamma$ have been derived in the literature, see \cite{hklw16} for discussion and recent results. The original version \cite{hklw02} is explicitly given by
\begin{equation}
\Gamma=\frac{2\gamma_2-\gamma_1^2}{24}\;\Big(\frac{\sigma C(F_{\mathrm{mid}})}{\alpha}\Big)^2+\frac{\rho\gamma_1}{4}\;\frac{\sigma C(F_{\mathrm{mid}})}{\alpha}+\frac{2-3\rho^2}{24}\,,
\end{equation}
where
\begin{equation}
\begin{split}
\gamma_1&=\frac{C'(F_{\mathrm{mid}})}{C(F_{\mathrm{mid}})}\\
&=\frac{\beta}{F_{\mathrm{mid}}}\;,
\end{split}
\end{equation}
and
\begin{equation}
\begin{split}
\gamma_2&=\frac{C''(F_{\mathrm{mid}})}{C(F_{\mathrm{mid}})}\\
&=-\frac{\beta(1-\beta)}{F_{\mathrm{mid}}^2}\;.
\end{split}
\end{equation}
The value $F_{\mathrm{mid}}$ denotes a conveniently chosen midpoint between $F$ and $K$ (such as the arithmetic average $(F+K)/2$).

It follows from \eqref{implVol} that the at the money volatility in the SABR model is given by
\begin{equation}
\begin{split}
\sigma^{\mathrm{ATM}}&=\sigma C(F)+O(\varepsilon)\\
&=\sigma F^\beta+O(\varepsilon),
\end{split}
\end{equation}
while the skew is
\begin{equation}
\begin{split}
\eta&=\sigma  C'(F)+O(\varepsilon)\\
&=\beta\sigma F^{\beta-1}+O(\varepsilon).
\end{split}
\end{equation}

\section{\label{greekSec}SABR greeks}

In this section we derive explicit expressions for the greeks in the SABR model, and in particular we obtain the modified delta and vega of \cite{b06}. To focus attention we use the normal Black-Scholes model as the basis for option pricing, and assume that the discounting interest rate is zero. We let $T$ denote the date on which the option expires and denote by $\tau=T-t$ the time to expiration.

Let $\cB$ denote the standard Black-Scholes pricing function in the normal model, i.e.
\begin{equation}
\cB(\tau,F,K,\sigma)=
\begin{cases}
\sigma\sqrt{\tau}\big(d_+ N(d_+)+N'(d_+)\big),\qquad \text{ for a call option,}\\
\sigma\sqrt{\tau}\big(d_- N(d_-)+N'(d_-)\big),\qquad \text{ for a put option,}
\end{cases}
\end{equation}
where $N(x)$ denotes the cumulative normal distribution, and where
\begin{equation}
d_\pm=\pm\;\frac{F-K}{\sigma\sqrt{\tau}}\;.
\end{equation}
Then the current time $t$ price $P_t$ of an option expiring at time $T$ under the SABR model is then given by
\begin{equation}
P_t=\cB(\tau,F_t,K,\sigma^{\mathrm{imp}}(\tau,F_t,K,\sigma_t)),
\end{equation}
where $\sigma^{\mathrm{imp}}$ is given by \eqref{implVol}. We should emphasize that this expression is only an approximation to the true SABR option price, to the degree to which the asymptotic implied formula \eqref{implVol} represents an accurate approximation to the true, analytically unknown expression for the SABR implied volatility (see \cite{hklw14} for an extensive discussion).

We decompose the Brownian motion $Z_t$ into $W_t$ and a Brownian motion $W^\perp_t$, independent of $W_t$: $Z_t=\rho W_t+\sqrt{1-\rho^2}\,W^\perp_t$. Then, $d\sigma_t$ can be written as a sum of $\rho\alpha/C(F_t)\,dF_t$ and a contribution $d\sigma^\perp_t$ uncorrelated with $dF_t$, namely $d\sigma^\perp_t=\alpha\sigma_t dW_t^\perp$. From Ito's lemma we obtain:
\begin{equation*}
\begin{split}
d\sigma^{\mathrm{imp}}_t&=-\frac{\d \sigma^{\mathrm{imp}}}{\d\tau}\,dt+\Big(\frac{\d \sigma^{\mathrm{imp}}}{\d F}+\frac{\d \sigma^{\mathrm{imp}}}{\d \sigma}\frac{\rho\alpha}{C(F_t)}\Big)dF_t+\frac{\d \sigma^{\mathrm{imp}}}{\d\sigma}\,d\sigma^\perp_t\\
&\quad+\frac12\,\sigma_t^2\Big(C(F_t)^2\,\frac{\d^2 \sigma^{\mathrm{imp}}}{\d^2 F}+2\rho C(F_t)\,\frac{\d^2 \sigma^{\mathrm{imp}}}{\d F\d\sigma}+\alpha^2\,\frac{\d^2 \sigma^{\mathrm{imp}}}{\d^2\sigma}\Big)dt.
\end{split}
\end{equation*}
This yields the following risk decomposition:
\begin{equation}\label{riskDec}
dP_t=\Big\{-\Theta_t+\frac12\,\sigma_t^2\big(C(F_t)^2\Gamma_t+2C(F_t)\mathrm{Vanna}_t+\alpha^2\mathrm{Volga}_t\big)\Big\}dt+\Delta^{\mathrm{mod}}_t dF_t+\mathrm{Vega}_t d\sigma^\perp_t,
\end{equation}
where the first and second order greeks are defined as follows:
\begin{equation}\label{bartDel}
\Delta^{\mathrm{mod}}_t=\frac{\d\cB}{\d F}+\frac{\d\cB}{\d\sigma}\Big(\frac{\d\sigma^{\mathrm{imp}}}{\d F}+\frac{\d\sigma^{\mathrm{imp}}}{\d\sigma}\frac{\rho\alpha}{C(F_t)}\Big)
\end{equation}
is the modified SABR delta,
\begin{equation}
\mathrm{Vega}_t=\frac{\d\cB}{\d\sigma}\frac{\d \sigma^{\mathrm{imp}}}{\d\sigma}
\end{equation}
is the SABR vega,
\begin{equation}
\Theta_t=\frac{\d\cB}{\d\tau}+\frac{\d\cB}{\d\sigma}\frac{\d\sigma^{\mathrm{imp}}}{\d\tau}
\end{equation}
is the SABR time decay,
\begin{equation}
\Gamma_t=\frac{\d^2\cB}{\d^2 F}+\frac{\d\cB}{\d\sigma}\frac{\d^2\sigma^{\mathrm{imp}}}{\d F^2}
\end{equation}
is the SABR gamma,
\begin{equation}
\mathrm{Vanna}_t=\frac{\d^2\cB}{\d F\d\sigma}+\frac{\d\cB}{\d\sigma}\frac{\d^2\sigma^{\mathrm{imp}}}{\d F\d\sigma}
\end{equation}
is the SABR vanna, and
\begin{equation}
\mathrm{Volga}_t=\frac{\d^2\cB}{\d^2 \sigma}+\frac{\d\cB}{\d\sigma}\frac{\d^2\sigma^{\mathrm{imp}}}{\d \sigma^2}
\end{equation}
is the SABR volga. Formula \eqref{riskDec} represents a risk decomposition of an option in terms of independent risk factors $dF$ and $d\sigma^\perp$, time decay, and second order greeks.

Alternatively, we can represent $W_t$ in terms of $Z_t$ and its independent complement $Z^\perp_t$ as $W_t=\rho Z_t+\sqrt{1-\rho^2}\,dZ^\perp_t$, and arrive at the following risk decomposition:
\begin{equation}\label{altRiskDec}
dP_t=\Big\{-\Theta_t+\frac12\,\sigma_t^2\big(C(F_t)^2\Gamma_t+2C(F_t)\mathrm{Vanna}_t+\alpha^2\mathrm{Volga}_t\big)\Big\}dt+\Delta_t dF^\perp_t+\mathrm{Vega}^{\mathrm{mod}}_t d\sigma_t.
\end{equation}
Here, the meaning of the greeks is as follows:
\begin{equation}
\Delta_t=\frac{\d\cB}{\d F}+\frac{\d\cB}{\d\sigma}\frac{\d\sigma^{\mathrm{imp}}}{\d F}
\end{equation}
is the standard SABR delta, and
\begin{equation}
\mathrm{Vega}^{\mathrm{mod}}_t=\frac{\d\cB}{\d\sigma}\frac{\d \sigma^{\mathrm{imp}}}{\d\sigma}+\Big(\frac{\d\cB}{\d\sigma}\frac{\d\sigma^{\mathrm{imp}}}{\d F}+\frac{\d\cB}{\d F}\Big)\frac{\rho C(F_t)}{\alpha}
\end{equation}
is the modified SABR vega. Formula \eqref{altRiskDec} is a decomposition of an option's risk in terms of an alternative basis of independent risk factors, namely $dF^\perp$ and $d\sigma$.

The two decompositions show that part of the option's volatility sensitivity can be viewed as a component of its delta or its vega, depending on risk management approach. We take the view that it should be allocated to the delta risk, as monitoring and executing delta hedges are generally easier than vega hedges.  Note also that the second order greeks do not contain any correlation dependent correction terms, and retain their form under both decompositions.

\section{\label{deltaSec}Robustness of the modified SABR delta}

We will now turn to the main point of this note and derive an explicit asymptotic expression for the modified SABR delta. Taking derivatives of \eqref{implVol} we find that, to within the leading order in $\varepsilon$,
\begin{equation}
\frac{\d\sigma^{\mathrm{imp}}}{\d F}=\frac{\alpha}{D(\zeta)}\Big\{1-\frac{\sigma^{\mathrm{imp}}}{\sigma C(F)I(\zeta)}\Big\}+O(\varepsilon),
\end{equation}
and
\begin{equation*}
\frac{\d\sigma^{\mathrm{imp}}}{\d \sigma}=\frac{\sigma^{\mathrm{imp}}\zeta}{\sigma D(\zeta) I(\zeta)}+O(\varepsilon).
\end{equation*}
In the following, in order not to overburden the formulas, we will be suppressing the terms $O(\varepsilon)$. It should be understood though that all formulas stated below are accurate to within $O(\varepsilon)$.

Now note that, for $\zeta$ small, we have
\begin{equation}
I(\zeta)=1-\rho\zeta+O(\zeta^2).
\end{equation}
As a consequence, the factor entering the modified delta \eqref{bartDel} can be written as
\begin{equation*}
\begin{split}
\frac{\d\sigma^{\mathrm{imp}}}{\d F}+\frac{\d\sigma^{\mathrm{imp}}}{\d \sigma}\frac{\rho\alpha}{C(F)}
&=\frac{\alpha}{D(\zeta)}\Big\{1-\frac{\sigma^{\mathrm{imp}}}{\sigma C(F)}\frac{1 -\rho\zeta}{I(\zeta)}\Big\}\\
&=\frac{\alpha}{D(\zeta)}\Big\{1-\frac{\sigma^{\mathrm{imp}}}{\sigma C(F)}+O(\zeta^2)\Big\}\\
&=\frac{\sigma^{\mathrm{imp}}}{F-K}\Big\{1-\frac{\sigma^{\mathrm{imp}}}{\sigma C(F)}+O(\zeta^2)\Big\}\\
&=\frac{\sigma^{\mathrm{imp}}}{\sigma C(F)}\frac{\sigma C(F)-\sigma^{\mathrm{imp}}}{F-K}+O(\zeta).
\end{split}
\end{equation*}
In the limit $K\to F$, we have $\sigma^{\mathrm{imp}}\to \sigma C(F)$, and hence
\begin{equation*}
\frac{\d\sigma^{\mathrm{imp}}}{\d F}+\frac{\d\sigma^{\mathrm{imp}}}{\d \sigma}\frac{\rho\alpha}{C(F)}
=\sigma C'(F)+O(F-K).
\end{equation*}

As a result of these calculations, the modified SABR delta is given by
\begin{equation}
\Delta^{\mathrm{mod}}=\frac{\d\cB}{\d F}+\frac{\d\cB}{\d\sigma}\,\eta+O(F-K),
\end{equation}
as claimed in the Introduction. Note that, to the leading order in the option moneyness, this expression is independent of the details of the backbone function $C(F)$, it only depends on the implied volatility for the strike $K$ and the skew $\eta$. Both of these quantities are market observable, and the calibrated model fits them. This explains the empirical observation made in \cite{b06} that the modified SABR delta is practically insensitive to the choice of the parameter $\beta$, once the remaining parameters have been optimized. In particular, the expression above shows that the modified delta of an at the money option, $K=F$, is independent of the choice of $\beta$.

This is to be contrasted with the behavior of the classic SABR delta. Indeed, we have
\begin{equation*}
\begin{split}
\frac{\d\sigma^{\mathrm{imp}}}{\d F}
&=\frac{\alpha}{D(\zeta)}\Big\{1-\frac{\sigma^{\mathrm{imp}}}{\sigma C(F)}\frac{1}{I(\zeta)}\Big\}\\
&=\frac{\alpha}{D(\zeta)}\Big\{1-\frac{\sigma^{\mathrm{imp}}}{\sigma C(F)}\,(1+\rho\zeta)+O(\zeta^2)\Big\}\\
&=\frac{\sigma^{\mathrm{imp}}}{\sigma C(F)}\Big\{\frac{\sigma C(F)-\sigma^{\mathrm{imp}}}{F-K}+\frac{\rho\sigma^{\mathrm{imp}}\zeta}{\sigma C(F)(F-K)}\Big\}+O(\zeta)\\
&=\sigma C'(F)+\frac{\rho\alpha}{C(F)}+O(F-K),
\end{split}
\end{equation*}
and therefore
\begin{equation}
\Delta=\frac{\d\cB}{\d F}+\frac{\d\cB}{\d\sigma}\Big(\eta+\frac{\rho\alpha}{C(F)}\Big)+O(F-K).
\end{equation}
In other words, the classic SABR delta, and thus the corresponding hedging strategy, depends on the choice of the backbone function $C(F)$.

\section{\label{empSec}Empirical analysis}

We will now discuss some numerical and empirical data supporting the arguments presented above. More evidence is described in \cite{b06}, \cite{am10}, \cite{hklw14} (for interest rate options), and in \cite{hw16}, \cite{cchp19} (for equity options).

\begin{figure}[H]
\scalebox{0.3}[0.24]{\includegraphics{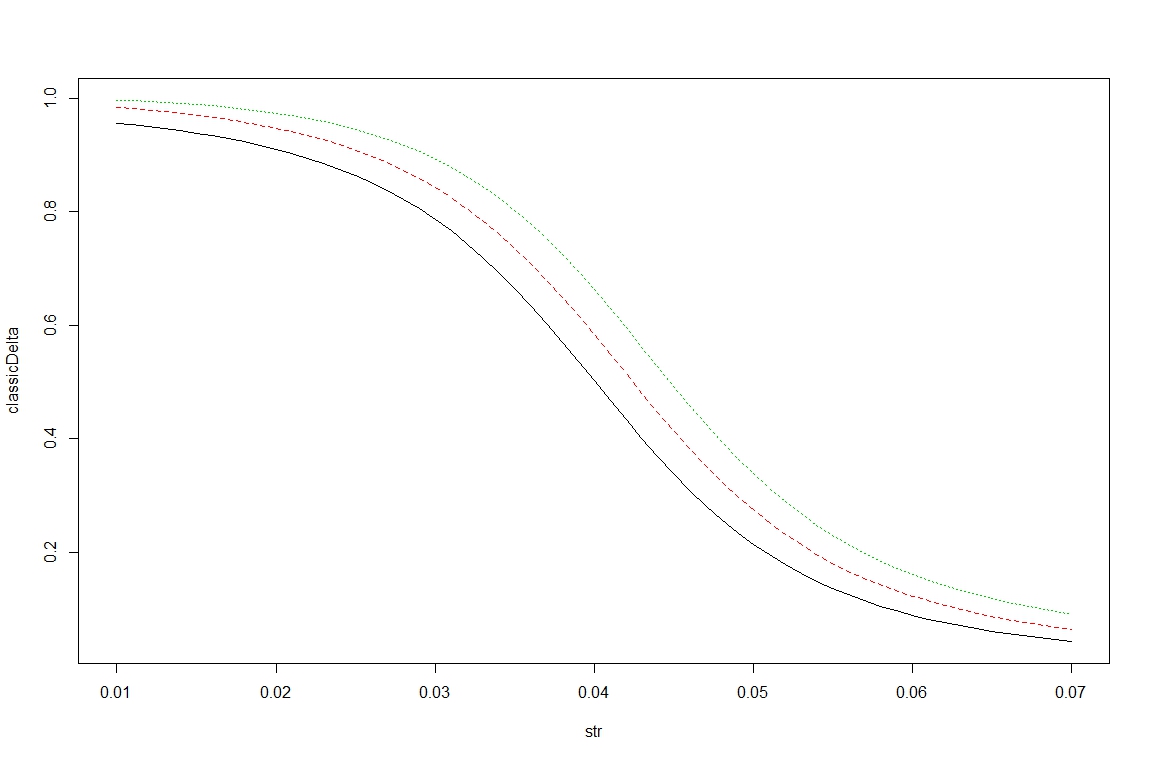}}
\caption{\label{betaClSabr} Classic SABR delta for different values of $\beta$.}
\end{figure}

Figure \ref{betaClSabr} shows the classic SABR delta corresponding to three different calibrations of the same smile curve: $\beta=0$ (black line), $\beta=0.5$ (red line), and $\beta=1$ (green line). For each of these choices of $\beta$, the three remaining SABR parameters are optimized to yield the best fit to the options prices corresponding to all available strikes $K$. Even though all three sets of parameters closely match the market smile, they lead to different delta hedges, especially for near the money strikes. Choosing the incorrect beta can lead to good fits of the smile, but may still produce relatively poor delta hedges. 

\begin{figure}[H]
\scalebox{0.3}[0.24]{\includegraphics{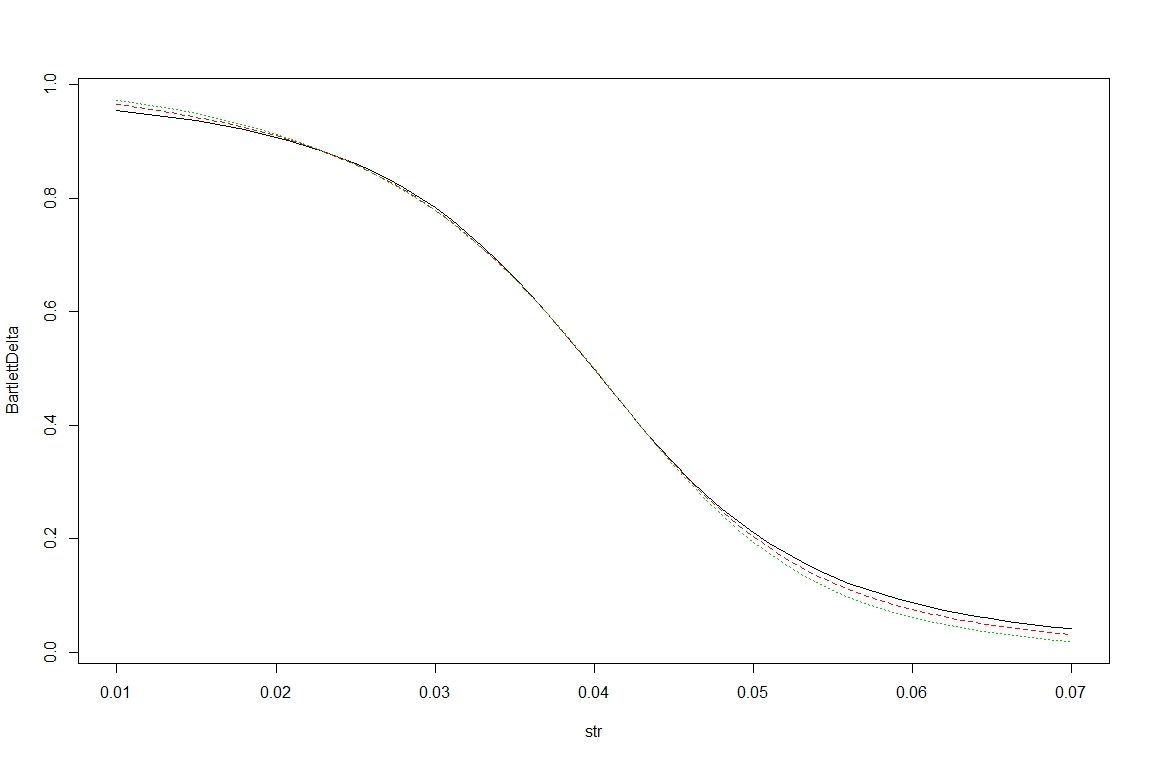}}
\caption{\label{betaBart} Bartlett's SABR delta for different values of $\beta$.}
\end{figure}

On the other hand, Figure \ref{betaBart} shows the modified deltas for the same three sets of parameters. Confirming the conclusions presented above, the modified SABR delta is nearly independent of $\beta$, even for way out of the money strikes. It depends mainly on the actual market smile, and not on how the smile is parameterized. Modified deltas tends to provide more robust hedges.

\begin{figure}[H]
\scalebox{1.7}[0.9]{\includegraphics{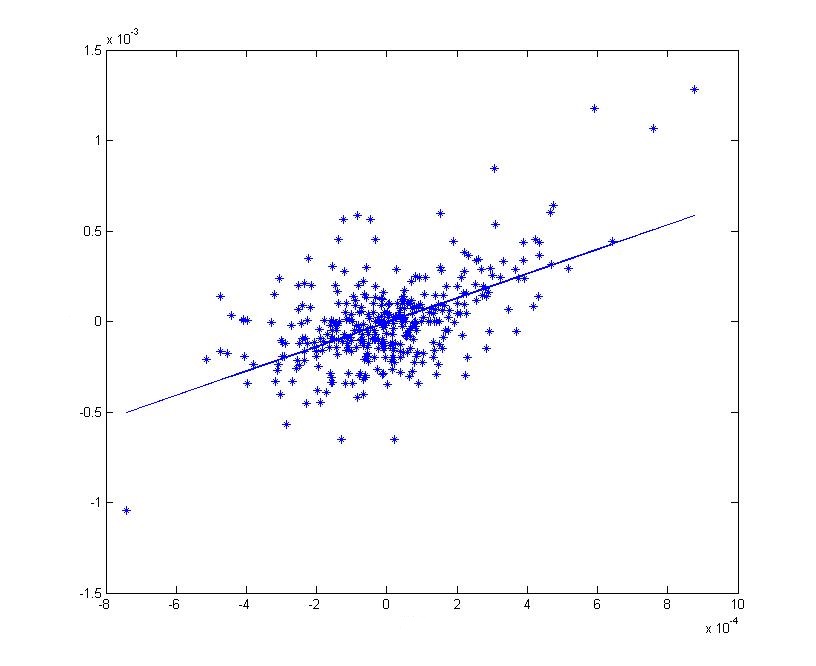}}
\caption{\label{1into10fig} Regression of $\delta\sigma$ against $\rho\alpha/F^\beta\,\delta F$ for the 1Y into 10Y swaption ($\beta=0.5$).}
\end{figure}

\begin{figure}[H]
\scalebox{1.7}[0.9]{\includegraphics{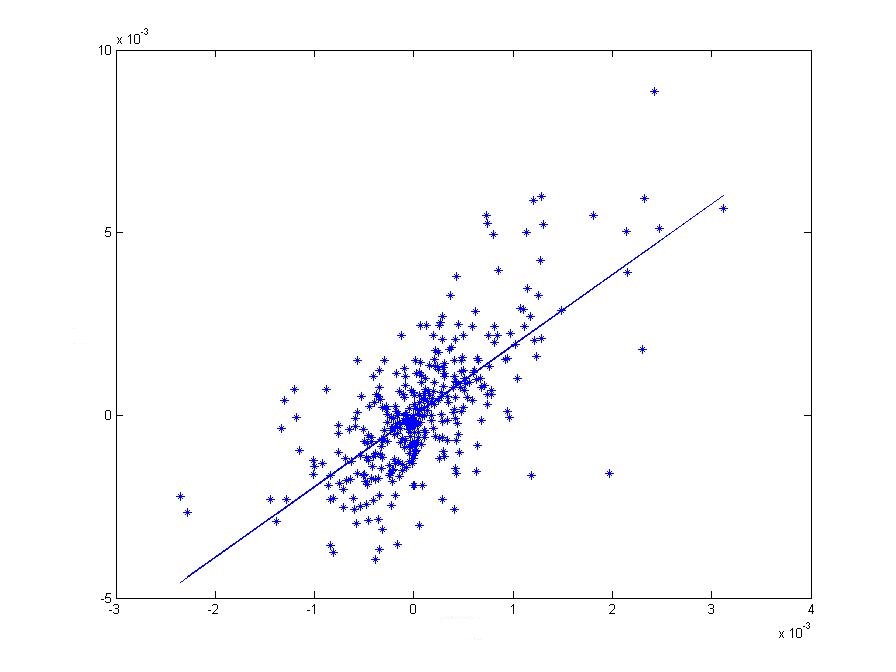}}
\caption{\label{5into5fig} Regression of $\delta\sigma$ against $\rho\alpha/F^\beta\,\delta F$ for the 5Y into 5Y swaption ($\beta=0.75$).}
\end{figure}

Figures \ref{1into10fig} and \ref{5into5fig} (both taken from \cite{am10}) present empirical data illustrating the historical relationship between the daily changes $\delta\sigma$ in the volatility parameters $\sigma$ and the daily changes in the forward swap rate $\delta F$, in the 1Y into 10Y and 5Y into 5Y swaption deltas, respectively. Specifically, the graphs represent the corresponding regressions of $\delta\sigma$ on $\rho\alpha/F^\beta\,\delta F$. The underlying data are historical closes from the period 2003 - 2010.

\end{document}